\documentstyle[proceedings,epsf]{crckapb} 

\makeatletter                                            
\@ifundefined{epsfbox}{\@input{epsf.sty}}{\relax}        
\def\plotone#1{\centering \leavevmode                    
\epsfxsize=\columnwidth \epsfbox{#1}}                    
\def\plotone_reduction#1#2{\centering \leavevmode        
\epsfxsize=#2\columnwidth \epsfbox{#1}}                  
\makeatother                                             

\begin{opening}
\title{COSMOLOGICAL ADAPTIVE MESH REFINEMENT}    

\author{M.L. NORMAN}
\institute{Astronomy Department \& NCSA\\
           University of Illinois, Urbana, IL 61801}
\author{G.L. BRYAN}
\institute{Princeton University Observatory\\
           Princeton, NJ 08544}                      

\end{opening}

\runningtitle{COSMOLOGICAL AMR}        

\begin{document}


\begin{abstract}
We describe a grid-based numerical method for 3D hydrodynamic cosmological
simulations which is adaptive in space and time and
combines the best features of higher order--accurate Godunov 
schemes for Eulerian hydrodynamics with adaptive
particle--mesh methods for collisionless particles. The basis for
our method is the structured adaptive mesh refinement (AMR) algorithm
of Berger \& Collela (1989), which we have extended to cosmological
hydro + N-body simulations. The resulting
{\em multiscale hybrid} method is a powerful alternative to particle-based
methods in current use. The choices we have made in constructing this
algorithm are discussed, and its performance on the Zeldovich pancake
test problem is given. We present a sample application of our
method to the problem of {\em first structure formation}. 
We have achieved a spatial dynamic range $L_{box}/\Delta x > 250,000$ in a 3D 
multispecies gas + dark matter calculation,
which is sufficient to resolve the formation
of primordial protostellar cloud cores starting from linear matter 
fluctuations in an expanding FRW universe.
\end{abstract}

\section{Introduction}

The inclusion of hydrodynamics and atomic and radiation processes into 
cosmological structure formation simulations is essential for many 
applications of interest, including galaxy formation, the structure
of the integalactic medium, formation and evolution of X-ray clusters,
and cosmic reionization. A key technical challenge in such simulations
is obtaining high mass and spatial resolution in collapsing structures
within large cosmological volumes.
Algorithms employing Lagrangian particles to represent dark matter and
gas have a natural advantage in this regard as they automatically
place resolution elements where they are needed. 
Such methods are now well developed and in widespread use. 
These include the gridless $TreeSPH$ method of Hernquist \& Katz (1989) and
Katz, Weinberg \& Hernquist (1996), as well as the grid-assisted
$P^3M-SPH$ method (Evrard 1988; Couchman, Thomas \& Pierce 1995). 
Parallel versions of these methods have recently been developed (Dav\'{e},
Dubinski \& Hernquist 1997; Pearce \& Couchman 1997) which, when
run on massively parallel computers, 
can integrate $O(10^7)$ particles in hydrodynamic simulations, and
as many as $10^9$ particles in pure dark matter simulations 
(Couchman, these proceedings).

\def\etal{et al.~}
We and other members of the Grand Challenge Cosmology Consortium ($GC^3$)
have explored Eulerian hydrodynamics methods as an alternative to SPH in 
cosmological simulations (Cen 1992; Anninos, Norman \& Clarke 1994;
Kang \etal 1994; Bryan \etal 1995; Gnedin 1995; Pen 1995). 
Provided grids can be constructed which achieve the
necessary dynamic range (a non-trivial issue, as we shall see), 
Eulerian methods have a number of distinct advantages over SPH.
These are: (1) {\em speed}: the use of logically regular data structures
avoid time-consuming nearest neighbor searches resulting in $\sim
10 \times$ higher update rate; (2) {\em noise}: fluid is represented as
a continuum, not discrete particles, eliminating Poisson noise; 
(3) {\em density sampling}: because of point (2), low density cells are
computed as accurately as high density cells at the same cost; 
density gradients spanning many orders of magnitude can be accurately
simulated with 3-4 cells per decade per dimension.
(4) {\em integral form}: integral conservation laws are straightforward
to implement for mass, momentum, energy and magnetic flux which
are numerically conservative to machine roundoff.
In addition to the above-mentioned advantages which are generic to
any Eulerian method, we add the following for higher order
Godunov methods such as PPM or TVD: (5) {\em shock capturing}: 
shocks are captured in 1-2 cells with correct entropy generation and
non-oscillatory shocks; (6) {\em upwind}: wave characteristics are
properly upwinded for higher fidelity and stability; (7) {\em low
dissipation}: the use of higher order-accurate interpolation results
in a very low numerical viscosity--important for angular momentum
conservation in protogalactic disks. In addition, radiative transfer
and MHD 
is most easily done on a grid. Implicit algorithms generate large
sparse matrix equations for which iterative and direct linear systems
solvers are available.

In reference to SPH, we mention
several disadvantages of traditional Eulerian methods: (1) {\em resolution}:
limited to the grid spacing $\Delta x$; (2) {\em invariance}: solutions
are not strictly translational and rotational invariant due to the
dependence of truncation errors on the relative velocity between
fluid and grid and grid orientation.   

In the references cited above, various gridding schemes have been 
explored to reduce truncation error in regions of high density
gradients, as invariably arise in structure formation simulations. 
Here, we describe a powerful method we have developed
based on the adaptive mesh refinement (AMR)
algorithm of Berger and Colella (1989).   
The paper is organized as follows. In Sec. 2 we briefly review the
elements of Berger's AMR. In Sec. 3 we describe the modifications
we have made to extend AMR to cosmological hydro+N-body simulations.
In Sec. 4 we test the method against several standard test problems
in numerical cosmology. In Sec. 5 we illustrate the power of AMR
in an application to the formation of the first baryonic objects
in a CDM-dominated cosmology. 
 
\section{Overview of Berger's AMR}

Adaptive mesh refinement (AMR) was developed by Berger and Oliger (1984)
to achieve high spatial and temporal resolution in regions of solutions
of hyperbolic partial differential equations where fixed grid methods
fail. Algorithmic refinements and an application to shock hydrodynamics
calculations were described in Berger \& Collela (1989). The hydrodynamic
portion of our method is based closely on this latter paper, and we refer 
the reader to it for details (see also paper by Klein et al., these
proceedings). 

\subsection{Grid Hierarchies}

\begin{figure}
\centerline{\epsfxsize 3.5in \epsfbox{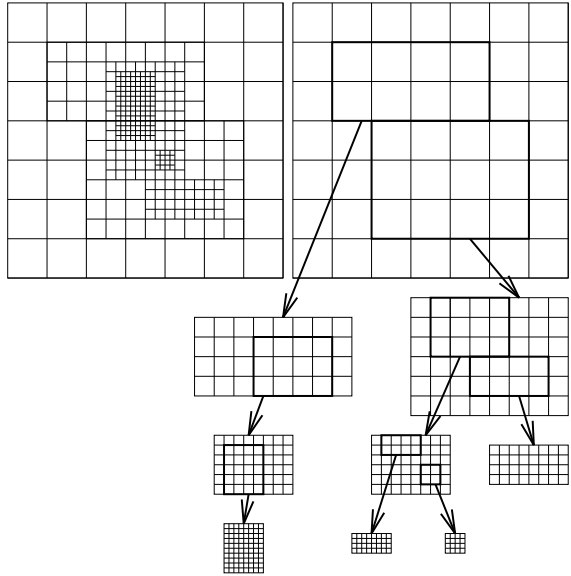}}
\caption{Illustration of an AMR grid hierarchy.}
\end{figure}

Unlike some mesh refinement schemes which move the mesh points
around ``rubber mesh"  (e.g., Dorfi \& Drury 1987) or subdivide individual
cells, resulting in an octree data structure (e.g., 
Adjerid, S. \& Flaherty 1998), Berger's AMR 
(also becoming known as {\em structured} AMR to differentiate it
from other flavors) utilizes an adaptive hierarchy of grid patches at various
levels of resolution. Each rectangular grid patch (hereafter, simply {\em grid})
covers some region of space in its {\em parent grid} needing higher
resolution, and may itself become the parent grid to an even higher
resolution {\em child grid}. A general implementation of AMR places
no restriction on the number of grids at a given level of refinement,
or the number of levels of refinement. The hierarchy of grids can be
thought of as a tree data structure, where each leaf is a grid. 
 
Each grid is evolved as a separate initial boundary value (IBV) problem.
Initial conditions are obtained by interpolation from the parent grid
when the grid is created. Boundary conditions are obtained either by
interpolation from the parent grid or by 
copies from abutting {\em sibling grids} at the same level.
To simplify interpolation, grids are constrained to be aligned with
their parent grid and not overlap their sibling grids. Grids
at level $\ell$ are refined by an integer factor $R_{\ell}
\geq 2$ relative to the grid at level $\ell -1$. 
Fig. 1 provides an illustration
of these concepts in a 2D, 4-level, $R_{\ell}=2$ grid hierarchy.  

\subsection{Algorithm}

The algorithm
which creates this grid hierarchy is {\em local} and {\em recursive}. 
It can be written in pseudocode as follows. Here \texttt{level} is
the level of refinement, and procedures are capitalized.

\begin{verbatim}
Integrate (level)
begin
  if "time for regridding" then Refine(level)
  CollectBoundaryValues(level)
  Evolve (level)
  if "level isn't finest existing" then begin
    for r = 1 to R do
      Integrate(level + 1)
    Update(level, level + 1)
  end
end
\end{verbatim}

Consider a typical calculation which is initialized on a single, 
coarsely resolved grid called the {\em root grid}, level  $\ell = 0$. 
The solution is integrated forward
in time with procedures \texttt{CollectBoundaryValues}
and \texttt{Evolve}. At each timestep, every cell is checked to
determine if it requires refinement--a process known as {\em selection}. 
Selection is based either on an estimate of the local truncation error,
or some simple threshold criterion, or some Boolean combination thereof.
If at least one cell is flagged, then procedure \texttt{Refine} 
is called. It does two things. First, flagged cells are {\em clustered}, 
and minimal rectangular boundaries of these clusters are determined using the
algorithm of Berger \& Rigoustous (1991). Second, one or more refined
$\ell = 1$ grids are allocated having these boundaries, and their intial
data is computed via interpolation from the root grid. 
Depending on circumstance, \texttt{Refine} may also deallocate
unneeded subgrids.
The root grid is then advanced one coarse timestep $\Delta t_0$.
The $\ell =1$ subgrids are integrated forward in time with 
smaller timesteps $\Delta t_1$ until they ``catch up"
with the parent grid. If the hyperbolic system is linear (e.g.,
advection), then $\Delta t_1 = \Delta t_0 /R$, and the refined
grids take R timesteps for each coarse grid timestep. However,
for nonlinear systems like gas dynamics, fine grid timesteps are 
determined by the Courant stability condition, 
and in general R+1 timesteps are needed to match times. 
This detail is not reflected in the pseudocode above. 
After the fine grids have been advanced to time $t + \Delta t_0$,
procedure \texttt{Update} is called. It does two things. First, it
injects the results of the fine grid calculation into the overlying
coarse grid cells through summation or interpolation. Second,
the values of coarse grid conserved quantities in cells adjacent to fine
grid boundaries are modified to reflect the difference between
the coarse and fine grid fluxes--a procedure known as {\em flux correction}.

The algorithm described above is recursive, and applies to any level 
in the grid hierarchy. 
In a multilevel grid hierarchy, the temporal
integration scheme is analogous to the ``W-cycle" in classic
elliptic multigrid. Grids are advanced from coarse to fine with
their individually determined timesteps. This order allows us to 
achieve second order accuracy in time on the entire hierarchy. 
This is accomplished by interpolating
the boundary values not only in space but in time as well
using time level n and n+1 values on the parent grid. 

\section{Cosmological AMR}

Cosmological hydrodynamic simulations require a robust fluid
solver capable of handling the extreme conditions of structure
formation, as well as a method for evolving collisionless particles
(dark matter, stars) subject to their self-consistent gravitational
field, the latter requiring a solution of the Poisson equation. 
We briefly describe these methods here. A more detailed
paper is in preparation (Bryan \& Norman 1998).

\subsection{Cosmological PPM}

Our fluid solver is based on the piecewise parabolic method (PPM)
of Collela \& Woodward (1984), suitably modified for cosmological
flows. The algorithm is completely described in Bryan \etal (1995),
so we merely state the essential points. PPM is a higher order-accurate
version of Godunov's method, featuring third order-accurate 
piecewise parabolic
monotonic interpolation and a nonlinear Riemann solver for shock capturing.
Multidimensional schemes are built up by directional splitting, where
the order of the 1D sweeps is permuted
a l\`{a} Strang (1968), resulting is a scheme which is formally 
second order-accurate
in space and time. For cosmology, the conservation laws for the fluid mass,
momentum, and energy density are written in comoving coordinates
for a FRW spacetime with metric scale factor a(t).
Both the gas internal energy equation and total (internal + kinetic) energy
equation are solved everywhere on the grid at all times. 
This {\em dual energy formulation} is adopted
in order to have a scheme that both produces the correct entropy
jump at strong shocks
{\em and} yields accurate pressures and temperatures in the
hypersonic parts of the flow. Both the conservation laws and the
Riemann solver must be modified to include gravity. In order
to maintain second order accuracy in time, the gravitational
potential $\phi$ is needed at the half time level $t^{n+\frac{1}{2}}$.
We use a predictor-corrector approach, wherein the particles are
advanced to $t^{n+\frac{1}{2}}$ using $\phi^n$, and then
$\phi^{n+\frac{1}{2}}$ is computed by solving the Poisson equation,
as described below.
We have implemented both Lagrange + Remap (LR) and Direct Eulerian (DE)
variants of our method, following the example of Collela \& Woodward (1984),
with comparable results. For AMR applications,
we use the DE version to simplify the flux correction step.

\subsection{Adding Collisionless Particles}

Adding collisionless particles to AMR presents two challenges, 
one physical (how do they interact with the fluid in the mesh), and
one algorithmic (how to add a new data structure.) 
A third challenge--how to compute the gravitational interaction 
between the two components in a consistent fashion--is described 
in the following section. 

Our method utilizes a single set of particles with comoving positions
$\vec{x_i}$, proper peculiar velocities $\vec{v_i}$ and masses $m_i$ 
(other characteristics
may be added as needed). There is a unique, one-to-one association
between particle $i$ and a grid $j$ at level $\ell ~G_{\ell , j}$ 
if that particle's position
lies within the grid's boundary but outside of any finer (child)
grid. In other words, a particle belongs to the finest grid which contains
it. We exploit this association by denoting that grid as the particle's
{\em home} grid and store all such particles in a list along with
the rest of the data connected to that grid. 
Note that a particle's home grid may change as it moves, requiring
redistribution of particles. This disadvantage is offset by a number of
factors, including (1) decreased search time for particle-grid interactions;
(2) improved data encapsulation; and (3) better parallelization characteristics
(see below).
This association is also very natural from a physical standpoint: because
the particles are indirectly connected to the solution on their home
grid, they tend to share the same timestep requirement.

The particles obey Newton's equations, which in the comoving frame are:
\begin{equation}
\frac{d\vec{x_i}}{dt} = \frac{1}{a} \vec{v_i},
\end{equation}
\begin{equation}
\frac{d\vec{v_i}}{dt} = - \frac{\dot{a}}{a} \vec{v_i}
 - \frac{1}{a}(\vec{\nabla} \phi)_i
\end{equation}
where the subscript $i$ in the last term in Eq. 2 means the 
gravitational acceleration
is evaluated at position $\vec{x_i}$. The gravitational potential
is computed from a solution of the Poisson equation, which takes
the form in comoving coordinates:
\begin{equation}
\nabla ^2 \phi =  \frac{4 \pi G}{a} (\rho - \bar{\rho}),
\end{equation}
\noindent
where $\rho$ is the local comoving mass density of gas and particles,
and $\bar{\rho}$ is its global average value.

These equations are finite-differenced and solved with the same
timestep as the grid, to reduce bookkeeping. Eq. (1) can be solved
relatively simply since only quantities local to the particle are
involved. We use a predictor-corrector scheme to properly time center
the RHS. Eq. (2) requires knowing the gravitational acceleration 
at postion $\vec{x_i}$. This is accomplished using the particle-mesh (PM)
method (e.g., Hockney \& Eastwood, 1980). In the first step, particle masses 
are assigned to the grid using second order-accurate TSC (Triangular
Shaped Cloud) interpolation. In the second step, the gridded particle density
is summed with the gas density, and then Eq. (3) is solved on the mesh
as described below. 
Finally, in the third step, gravitational accelerations are interpolated
to the particle positions using TSC interpolation, and particle velocities
are updated.

\subsection{Solving the Poission Equation}

The description above glossed over an important detail. Namely, that
the gravitational field cannot be solved grid by grid, but must have
knowledge of the mass distribution on the entire grid hierarchy.
However, we wish to use the high spatial resolution afforded by
the refined grids to compute accurate accelerations.
This is accomplished using a method similar to that set out in Couchman
(1991). The basic idea is to represent the acceleration on a particle
as the sum of contributions from each level of the hierarchy less than
or equal to the level of its home grid:

\begin{equation}
\vec{a_i} = \sum_{\ell ' = 0}^{\ell} \vec{a_i}(\ell ')
\end{equation}

\noindent
where the partial accelerations $\vec{a_i}(\ell)$ are computed symbolically
as follows:

\begin{equation}
\{\vec{x_i}, m_i\}_{G_{\ell}} \stackrel{TSC}{\longrightarrow}
\rho_{\ell}(\vec{x}_{G_{\ell}}) \stackrel{FFT}{\longrightarrow} 
\hat{\rho_{\ell}}(\vec{k}),
\end{equation}
\begin{equation}
\hat{\phi}_{\ell}(\vec{k})=\hat{\rho_{\ell}}(\vec{k})
\mathcal{G}\mathit{_{\ell}(\vec{k})}
\longrightarrow \hat{a_{\ell}}(\vec{k}),
\end{equation}
\begin{equation}
\hat{a_{\ell}}(\vec{k}) \stackrel{3FFTs}{\longrightarrow}
\vec{a}_{\ell}(\vec{x}_{G_{\ell}}) \stackrel{TSC}{\longrightarrow} 
\vec{a}_{\ell}(\vec{x}_i).
\end{equation}
In the first step, the mass of {\em all} particles whose positions
lie inside the boundaries of grid $G_{\ell}$ is assigned to the
grid using TSC interpolation, and its Fourier transform is computed.
In the second step the Poisson equation is solved in Fourier space
using a shaped force law $\mathcal{G}\mathit{_{\ell}(\vec{k})}$ designed to 
reproduce a $\frac{1}{r}$ potential when summed over all levels.
Accelerations on the grid are computed in Fourier space, and
then transformed back into real space in step 3. This requires
three FFTs---one for each component. Finally, the acceleration
on particle $i$ due to $G_{\ell}$ is computed using TSC interpolation.
The acceleration on the fluid is computed
by treating each cell as a particle with its
position given by the center of mass (as determined by a
tri-linear interpolation).  The assignment of mass and
interpolation of acceleration is done with the same TSC scheme as
used for the particles.

\subsection{Refinement Factor and Criteria}

In our implementation, the refinement factor $R_{\ell}$ can have any
integer value, and can be different on different levels.
However, we have found through experimentation that $R_{\ell}=R=2$ 
is optimal for cosmological simulations where the gravitational 
dynamics is dominated by dark matter. 
Since dark matter is represented by
a fixed number of particles, the use of higher refinement factors
$R \geq 3$ refines the gas grid to the point where each dark matter 
particle becomes an accretion center. The choice R=2 maintains
commensurate mass resolution in both gas and dark matter.

In order to reduce Poisson noise, we initialize a calculation
with one particle per cell. We flag a cell for refinement when
the baryonic mass has increased by a factor of four over its
initial value. Immediately after refinement, refined cells in 3D
have one half their initial mass and typically contain
zero or one particle. Thus,
in a collapsing stucture, the ratio of baryonic to dark matter
mass will vary between one half and four times its initial
value.

\subsection{Implementation}

Our implementation features arbitrary refinement factors
$R_{\ell}$, number of grids $N_g$, 
grid shapes ($n_x \neq n_y \neq n_z$),
and grid hierarchy depth $\ell_{max}$. 
We also have the option, not yet exercised, of calling different
physics solvers on different levels.
The AMR driver is written in object-oriented C++ to simplify
logic and memory management. 
The grid objects encapsulate the grid data (field
and particle) as well as numerous grid methods (e.g., \texttt{Evolve}).
The floating-point intensive methods are implemented in F77 for
the sake of computational efficiency. 
C wrappers interface the C++ and F77 code.
Our production version is
a shared memory, loop parallel version, where all grids at a 
given level are executed in parallel. A distributed memory
version, wherein the root grid is domain decomposed,
is under development.

\section{Tests}

\begin{figure}
\centerline{\epsfxsize 4.5in \epsfbox{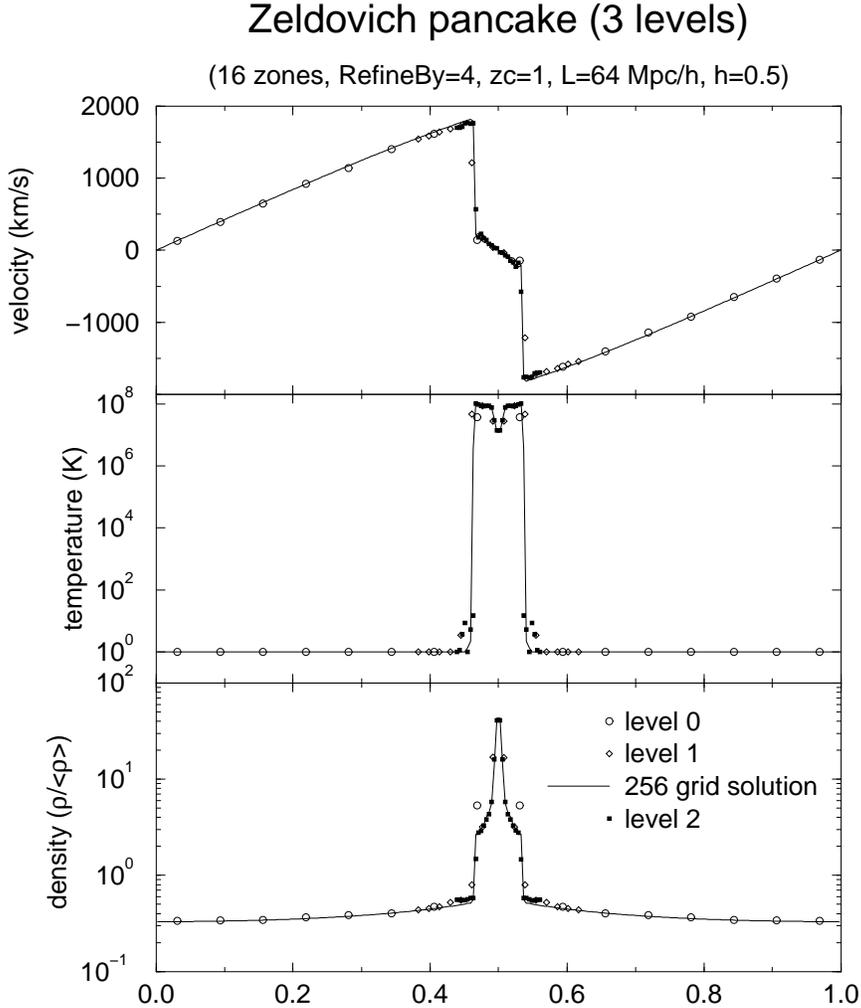}}
\caption{1D Zeldovich pancake test problem. 3-level solution (symbols)
is overlaid on 256 zone uniform grid solution (solid line).}
\end{figure}

We have extensively tested our code against a variety of hydrodynamic
and hybrid (hydro + N-body) test problems. These include (1) linear
waves and shock waves entering and exiting static subgrids in 1D and
2D at various angles; 
(2) a stationary shock wave in a static 3D subgrid ``Shock in a 
Box" (Anninos, Norman \& Clarke 1994); (3) Sod shock tube;
(4) 1D pressureless collapse; (5) 1D Zeldovich pancake;
(6) 2D pure baryonic CDM model (Bryan \etal 1995); (7) 3D adiabatic
X-ray cluster ``Santa Barbara cluster"
(Frenk \etal 1998); and (8) 3D self-similar infall
solution (Bertschinger 1985). In tests, (3-7) AMR results were
compared with unigrid results up to grid sizes of $512^3$. 
In all cases the algorithm performs well, reproducing the reference
solutions with a minimum of artifacts at coarse/fine grid interfaces.
The use of upwind, second-order accurate fluxes in space and time
is found to be essential in minimizing reflections. Details are
provided in Bryan \& Norman (1998).

As an example, Fig. 2 shows the code's performance on the 1D
Zeldovich pancake test problem. The parameters for this problem,
which includes gas, dark matter, gravity, and cosmic expansion,
are given in Bryan \etal 1995. Here we compare a 3-level AMR
calculation with a 256 zone uniform grid calculation. The root
grid has 16 zones, and the refinement factor R=4 (in 1D problems
one has more latitude with R.) Thus, the
AMR calculation has the same 256 zone resolution on the finest
grid $\ell = 2$. We can see that the AMR algorithm places the 
refinements only in the central high density pancake. The
solutions on the three levels of refinement match smoothly
in the supersonic infalling envelope. The density maximum and
the local temperature minimum at the midplane agree with the
uniform grid results exactly. The shock waves at x=.45 and .55
are captured in two cells without post-shock oscillations. 
The temperature field--the most difficult to get right--exhibits
a small bump upstream of the shock front. This is
caused by our dual energy formulation, and marks the location
where we switch from internal energy to total energy as
a basis for computing pressures and temperatures. 
The bump has no dynamical effect as ram pressure dominates
by many orders of magnitude ahead of the shock.

\section{A Sample Application: First Structure Formation}

We have applied our code to the formation of X-ray clusters (Bryan \&
Norman 1997), galaxy formation (Kepner \& Bryan 1998), and the
formation of the first baryonic structures in a CDM-dominated universe
(Abel, Bryan \& Norman 1998). In Fig. 3 we show a result from the latter
calculation to illustrate the capabilities of our method.

\begin{figure}
\centerline{\epsfxsize 5in \epsfbox{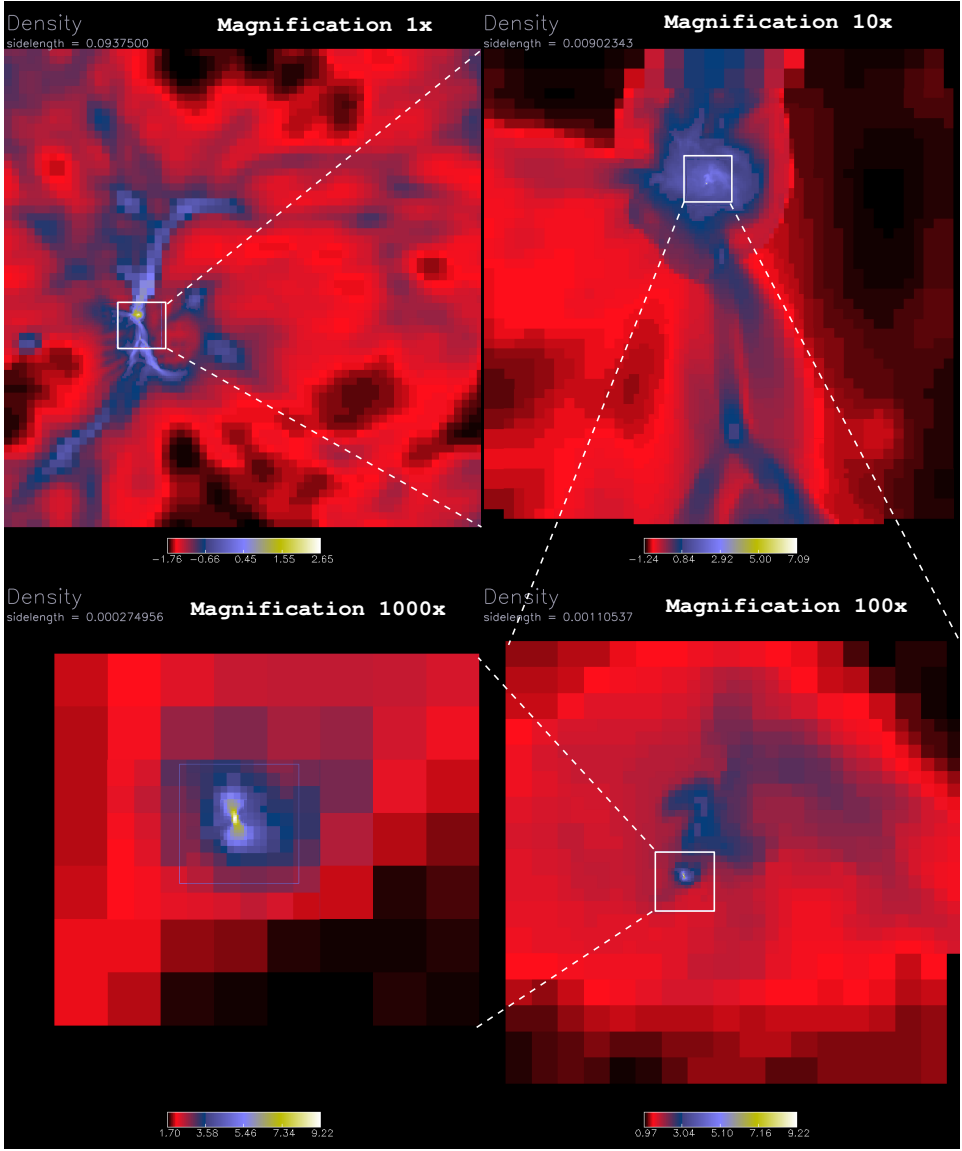}}
\caption{AMR simulation of first structure formation. Plotted
at four magnifications is the logarithm of the baryonic overdensity 
on a slice through the densest structure.} 
\end{figure}

Our ultimate goal is to simulate the formation of the first stars in the
universe (Population III stars) starting from cosmological initial
conditions. To resolve the protostellar cloud cores which form these
stars requires a spatial dynamic range of at least five
orders of magnitude in 3D. The primary coolant in zero
metallicity gas for $T < 10^4 K$ is $H_2$ line cooling 
(e.g., Tegmark \etal 1997). $H_2$
forms in the gas phase via nonequilibrium processes (McDowell 1961;
Saslaw \& Zipoy 1967). Thus, in addition to gas and dark matter, we 
also solve a 9-species nonequilibrium
chemical network for $H, H^+, He, He^+, He^{++},$ 
$ H^-, H_2^+, H_2$ and
$e^-$ in every cell at every level of the hierarchy. 
This requires adding nine new field variables--one for each 
species--and solving
the stiff reactive advection equations for the coupled network. 
The physical model is described in Abel \etal (1997). The numerical
algorithms for cosmological reactive flow are 
provided in Anninos \etal (1997).

We simulate a standard CDM model in a periodic,
comoving volume 128 kpc on a side. The parameters are: 
$\Omega _0 = 1, \Omega _b=0.06, \sigma _8=0.7, H_0=50$ ~km/s/Mpc.
The starting redshift is z=100. 
The intial conditions are realized on a 3-level static nested grid hierarchy
with a root grid resolution of $64^3$ cells. The subgrids are
centered on the most massive peak which develops. 
This location is determined by running the simulation once at low
resolution. The $\ell = 2$ grid is made large enough so that it initially
contains all the mass that eventually ends up in the condensed halo.
The initial $\ell = 2$ mass resolution in the gas (dark matter) 
is $0.5 (8) M_{\odot}$, respectively.

As the calculation proceeds, the AMR algorithm generates many more
grids at higher levels of refinement up to a preset limit of 13
levels total. In addition to the overdensity refinement criterion
described above, we also require that the numerical Jeans criterion
is satisfied everywhere (Truelove \etal 1997; Klein \etal, these proceedings.)
Nonlinear structures form in the gas by $z \approx 30$. By $z \approx 20$
sufficient $H_2$ has formed in the center of the virialized halo that
a cooling flow ensues (Abel \etal 1998). Because of the range of
mass scales which go nonlinear nearly simultaneously at these epochs,
the halo is quite lumpy as structure builds up hierarchically. By
$z=18$ (Fig. 3), a highly concentrated structure has formed  
through the collision of two rapidly cooling blobs. A collapsing,
primordial protostellar core forms as a result. The core collapses
to higher densities, eventually reaching our resolution
limit. The proper cell size on the 
$\ell = 12$ grid is $\Delta x_{12} = 0.025$ pc $\sim 5000$ AU. The
overall dynamic range achieved is $64*2^{12}=262,144.$

We stop the calculation when the Jeans criterion is violated
on the $\ell = 12$ grid. This occurs in the densest cell, which
has reached a baryonic overdensity of almost $10^{10}$ and
a proper number density of more than $10^{8}$ ~cm$^{-3}$. At these
densities, three body reactions become important, which we have
not included in our model. At the end of the calculation, we
have formed a parsec-sized collapsing primordial core with about
$500 M_{\odot}$ of material, roughly equal parts gas and dark matter.
The further evolution of this cloud, including the important
question of fragmentation, is being studied with a separate,
smaller scale AMR simulation in progress. 

\bigskip
\noindent
\textbf{Acknowledgements}
We thank our collaborators Tom Abel and Peter Anninos for
joint work cited here, as well as Sir Martin Rees for drawing
our attention to the problem of first structure formation.
This work was carried out under the auspices of the Grand Challenge
Cosmology Consortium ($GC^3$), with partial funding provided by
NSF grant ASC-9318185 and NASA grant NAG5-3923. Simulations
were carried out using the SGI Origin2000 at the National Center
for Supercomputing Applications, University of Illinois, Urbana-Champaign.

\bigskip
\noindent
\def\apj{Ap. J.}
\def\apjl{Ap. J. (Lett.)}
\def\apjs{Ap. J. Suppl.}
\def\mnras{M.N.R.A.S.}
\def\newa{New A.}
\def\etal{{\it et al. }}
\def\araa{Ann. Rev. A. \& A.}

\end{document}